\documentclass[10pt,twocolumn,twoside]{IEEEtran}

\usepackage{amsfonts}
\usepackage{amssymb}
\usepackage{cite}
\usepackage{graphicx}
\usepackage[cmex10]{amsmath}
\usepackage[]{algorithm}
\usepackage{algorithmic}
\usepackage{float}
\usepackage{subfigure}

\ifCLASSINFOpdf
\else
\fi
\begin{document}
%
\title{Deep Learning Based Fast Multiuser Detection for Massive Machine-Type Communication}
%
%
%

\author{Yanna~Bai, Bo~Ai,~\IEEEmembership{Senior Member,~IEEE}, and~Wei~Chen*,~\IEEEmembership{Senior Member,~IEEE}
\thanks{Yanna Bai, Bo Ai and Wei Chen are with the State Key Lab of Rail Traffic Control and Safety,
Beijing Jiaotong University, Beijing, China (e-mail: {16125001,boai,mmni,weich}@bjtu.edu.cn).}
\thanks{Corresponding author: Wei Chen.}
}

\maketitle

\begin{abstract}
Massive machine-type communication (MTC) with sporadically transmitted small packets and low data rate requires new designs on the PHY and MAC layer with light transmission overhead. Compressive sensing based multiuser detection (CS-MUD) is designed to detect active users through random access with low overhead by exploiting sparsity, i.e., the nature of sporadic transmissions in MTC. However, the high computational complexity of conventional sparse reconstruction algorithms prohibits the implementation of CS-MUD in real communication systems. To overcome this drawback, in this paper, we propose a fast Deep learning based approach for CS-MUD in massive MTC systems. In particular, a novel block restrictive activation nonlinear unit, is proposed to capture the block sparse structure in wide-band wireless communication systems (or multi-antenna systems). Our simulation results show that the proposed approach outperforms various existing algorithms for CS-MUD and allows for ten-fold decrease of the computing time.
\end{abstract}


\IEEEpeerreviewmaketitle

\section{Introduction}
%
%
%
%


\IEEEPARstart{R}{ecent} years observe a growing interest in massive machine-type communication (MTC) owing to the rapid development of Internet of Things and 5G~\cite{7565189}. In a typical MTC communication scene, a massive number of nodes sporadically transmit small packets with a low data rate, which is quite different to current cellular systems that are designed to support high data rates and reliable connections of a small number of users per cell. Communication overhead takes up a larger portion of resources in the MTC scene, and thus more efficient access methods are needed. One potential approach to reduce the communication overhead is to avoid or reduce control signaling overhead regarding the activity of devices before transmission.


\par
An important issue in massive MTC is access congestion owing to the large number of MTC devices. In~\cite{lien2011toward}, four approaches, i.e., backoff-based scheme, access class barring based scheme, separating random access channel (RACH) resources and dyanamic allocation of RACH resources, are introduced to deal with access congestion. However, those approaches does not effectively reduce signaling overhead. In massive MTC with sporadic communication, a compressive sensing (CS) based multiuser detection (MUD)~\cite{6125356,bockelmann2013compressive} is proposed to joint detect user activity and data with a known channel state information (CSI), which reduces communication overhead by eliminating control signaling. In practical systems where CSI is unknown, a CS based joint activity and channel detection method is proposed in~\cite{6629742}, where each node is assigned a unique pilot sequence for channel estimation. The cost brought by the CS-MUD approach is the effort in solving a sparse estimation problem, which requires iterative algorithms, e.g., orthogonal matching pursuit (OMP)~\cite{6145475}. However, these iterative algorithms are designed and optimized for achieving a higher accuracy and/or theoretically guaranteed convergence and fail to consider time constraints. Applying iterative steps in traditional sparse estimation algorithms until achieving convergence would increases communication latency (especially when the number of nodes is large), which is critical in some applications. It then naturally begs the question: can we reduce the signaling overhead for MTC without significantly sacrifice on the latency.

\par
In this paper, we propose a fast Deep Learning (DL) based approach for MUD in massive MTC systems. As one of the most highly sought-after skills in technology, DL~\cite{lecun2015deep} has been applied to various fields including computer vision~\cite{7780459}, speech recognition~\cite{6296526}, and language translation~\cite{sutskever2014sequence} and got great success. As a branch of machine learning, DL, which usually refers to deep neural networks (DNNs), consumes a large amount of training data to learn parameters in a neural network. When the neural network has a sufficient number of hidden units, it can approximate a large class of piecewise smooth functions~\cite{hornik1989multilayer}. Although the training process of the proposed DL based approach is time consuming, it can be conducted off-line with synthetically generated data. For the inference task, i.e., MUD, the computing complexity of the trained DNN is low, as it only involves a number of vector-matrix multiplications/summations and element-wise nonlinear operations.

\par
In addition, for wide-band wireless communication systems or multi-antenna systems, the transmitted signal arrives at the receiver with multiple paths or multiple links, which leads to a block sparse CSI vector. Capitalizing on the block sparse structure, we further propose a novel block restrictive activation nonlinear unit, which is distinct to existing activation functions in DNNs~\cite{wang2016learning,xin2016maximal}. Experiments demonstrate the efficiency and effectiveness of the proposed block-restrict neural network (BRNN) in compared with existing methods. 



\section{Background}
\subsection{System Description}
\begin{figure}[!tb]%
\centering%
\includegraphics[width=0.38\textwidth]{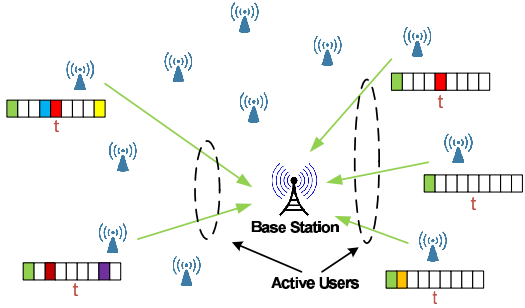}%
\DeclareGraphicsExtensions. \caption{A star-topology for the massive MTC scenario.} \label{fig:cs_sleep}
\end{figure}%

In this paper, we consider a massive MTC scenario where multiple devices communicate with a base station (BS), as shown in Fig.~\ref{fig:cs_sleep}. Without loss of generality, only $n$ devices out of $K$ devices have data to be transmitted to the BS in one frame. For simplicity, we assume that all frames are received synchronous at the BS. Each user is assigned a unique pilot sequence $\mathbf{s}_k$ ($k=1,\ldots,K$) with the length $N_\mathrm{s}$ for channel estimation. Each symbol of the pilot sequence is chosen from the modulation alphabet $\mathcal{A}$. The channel vector $\mathbf{h}_k$ of user $k$ is denoted by $\mathbf{h}_k\in\mathbb{C}^{L}$, where $\mathbb{C}$ denotes the set of complex numbers. The convolution of the channel vector and the pilot sequence $\mathbf{s}_k$ can be expressed as the matrix multiplication by rewriting the convolution matrix of the transmitted pilot sequence of user $k$ as

\[\hat{\mathbf{S}}_k=
\begin{bmatrix}
s_{k,1}	&0				&\cdots		\\
s_{k,2}	& s_{k,1}		&		\\
\vdots			&\vdots			&\ddots		\\
s_{k,N_\mathrm{s}}	&s_{k,N_\mathrm{s}-1}	&		\\
0			&s_{k,N_\mathrm{s}}	&		\\
0			&0				&\ddots
\end{bmatrix}\in \mathbb{C}^{N},\]
where $ N=(N_\mathrm{s}+L-1)\times L$. Then the signal received at the BS is given by
\begin{equation}\label{eq:CS_measurement}
\mathbf{y}=\sum\limits_{k=1}^{K}a_k\hat{\mathbf{S}}_k\mathbf{h}_k+\mathbf{n},
\end{equation}
where $\mathbf{n}$ denotes additive white Gaussian noise, and $a_k\in \{0,1\}$ denotes the activity of user $k$. $a_k=1$ and $a_k=0$ indicate active user and silent user, respectively.

\par
Now we construct the pilot matrix of all user as $\hat{\mathbf{S}}=[\hat{\mathbf{S}}_1,\ldots,\hat{\mathbf{S}}_K]\in \mathbb{C}^{(N_\mathrm{s}+L-1)\times K L}$, the channel vector of all user as $\mathbf{h}=[\mathbf{h}_1^T,\ldots,\mathbf{h}_K^T]^T\in \mathbb{C}^{K L}$, and the user activity matrix as $\mathbf{A}=\text{diag}(a_1\mathbf{I},\ldots,a_K\mathbf{I})\in \mathbb{R}^{K L\times K L}$, where $\text{diag}(\cdot)$ denotes the transformation of a vector into a diagonal matrix. Then the equation (\ref{eq:CS_measurement}) can be reexpressed as
\begin{equation}\label{eq:model}
\mathbf{y}=\hat{\mathbf{S}}\mathbf{A}\mathbf{h}+\mathbf{n}=\hat{\mathbf{S}}\mathbf{x}+\mathbf{n},
\end{equation}
where $\mathbf{x}=\mathbf{A}\mathbf{h}$ is a block sparse vector with $n$ nonzero blocks corresponding to active users. By reconstructing $\mathbf{x}$ from $\mathbf{y}$, we simultaneously detect active users and estimate their channel. Therefore, the problem boils down to solve the following optimization problem
\begin{equation}\label{eq:IHT}
\min_{\mathbf{x}}\ \left\|\mathbf{y}-\hat{\mathbf{S}}\mathbf{x}\right\|_2^2,\ \
\text{s.t.}\
\left\|\mathbf{x}\right\|_0\leq nL,
\end{equation}
where $\|\cdot\|_0$ denotes the $\ell_0$ norm that counts the number of nonzero elements. Note that the optimization problem in (\ref{eq:IHT}) is NP-hard, and popular approximations with varying degrees of computational overhead include convex relaxation methods~\cite{kukreja2006least} and iterative algorithms~\cite{blumensath2009iterative}.

\subsection{Deep Neural Network}
From the perspective of DL, the process that solves the optimization problem in (\ref{eq:IHT}) could be seen as a black box, which is expressed as a function
\begin{equation}\label{eq:DL_blackbox}
g(\mathbf{y},\hat{\mathbf{S}},\boldsymbol{\theta})=\arg\min_{\left\|\mathbf{x}\right\|_0\leq nL}\left\|\mathbf{y}-\hat{\mathbf{S}}\mathbf{x}\right\|_2^2,
\end{equation}
where $\boldsymbol{\theta}$ denotes a set of parameters. Given a set of training examples $\mathcal{D}=\{\mathbf{x}^{(i)},\mathbf{y}^{(i)}\}_i$, a DNN is learned to map each input $\mathbf{y}^{(i)}$ to a desired outcome by several successive layers of linear transformation interleaved with element-wise non-linear transforms. For an ordinary feedforward neural network (FNN), the $t$th layer can be expressed as
\begin{equation}\label{eq:DNN}
\mathbf{x}^{t+1}=f(\mathbf{W}^{t}\mathbf{x}^{t}+\mathbf{b}^{t}),
\end{equation}
where the weight matrix $\mathbf{W}^{t}$ and the bias vector $\mathbf{b}^{t}$ are parameters to be learned, and $f(\cdot)$ denotes some non-linear operator, e.g., rectilinear units (RELU).

\par
Various DNN designs for the sparsity enforcing problem as (\ref{eq:IHT}) have been proposed in literature. For example, in comparison to the ordinary FNN layers as defined in (\ref{eq:DNN}), a learned iterative shrinkage and thresholding algorithm (LISTA) is proposed in~\cite{gregor2010learning}, where different layers share same parameters, i.e., $\mathbf{W}$ and $\mathbf{b}$. Furthermore, the nonlinear unit $f[\cdot]$ adopted in LISTA is the element-wise soft-thresholding function $f[x]=sign(x)max\{0,|x|-\delta\}$, where $\delta$ is the shrinkage parameter. An IHT-net is proposed in~\cite{wang2016learning}, which is the same as LISTA except for using a hard thresholding function $f[x]=sign\{x\}max\{0,|x|\}$ as the nonlinear unit. While both LISTA and IHT-net use shared weights among layers, authors in~\cite{xin2016maximal} propose to use ordinary FNN where layers do not have shared weights, and incorporate batch normalization~\cite{ioffe2015batch} and residual connection~\cite{7780459} to reasonably initialize the neural network and to prevent vanishing/exploding gradients, respectively.

\section{Proposed Approach}
\subsection{Network Structure}
\begin{figure*}[!t]%
\centering%
\includegraphics[width=0.8\textwidth]{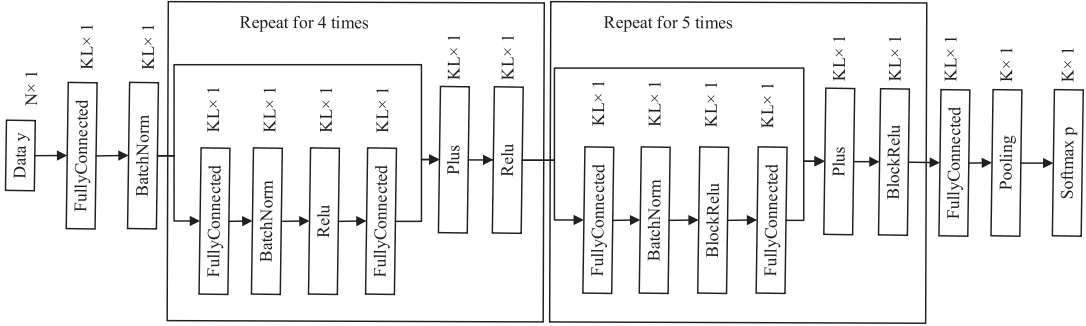}%
\DeclareGraphicsExtensions. \caption{The structure of the proposed BRNN. }
\label{fig:network}
\end{figure*}%
\par
The most straightforward DNN design for tackling a regression problem as in (\ref{eq:IHT}) is to map the received signal $\mathbf{y}$ to some outcome $\mathbf{x}$. In view of the fact that a sparse $\mathbf{x}$ can be obtained by the least square estimator given its support, it is would be more capacity-efficient to use a DNN to approximate the mapping from $\mathbf{y}$ to the support of $\mathbf{x}$. Therefore, we consider to use DNN for
detecting active users, which leads to a multi-label classification problem.

\par
Furthermore, in wideband wireless communication systems, $\mathbf{x}$ becomes a block sparse vector\footnote{For narrowband systems ($L=1$) with multiple antennas at the BS, $\mathbf{x}$ is also a block sparse vector, where the block length is the number of antennas.} with the block size $L$. It would be beneficial to incorporate this prior information into the structure of the designed DNN. Here, we propose to use a new block activation unit
\begin{equation}\label{eq:block_relu}
f(x_1,\ldots,x_L)=\text{sign}(\max\{0,x_1,\ldots,x_L\})\cdot[x_1,\ldots,x_L],
\end{equation}
where a block of elements are jointly activated if one element is greater than zero. Here $\text{sign}(\cdot)$ denotes the sign function. Batch normalization is added for reasonable initialization and residual connection is used to prevent vanishing/exploding gradients. Furthermore, we adopt a pooling layer before the last softmax layer to force the output of the network to indicate the active users. RELU is employed in the first a few layers, while the block activation unit is used in the remaining layers. The last layer of BRNN employs the softmax cross entropy loss function. This proposed network is named as BRNN, and its structure is illustrated in Fig.~\ref{fig:network}.

\subsection{Training Data Generation}
A common issue in applying DL to wireless communication is the difficulty of collecting massive real data. Fortunately, for solving the optimization problem in (\ref{eq:IHT}) by a DNN, we could use synthetically generated data for training. In specific, to obtain each training data, we first generate a random noise vector $\mathbf{n}$ and a random block sparse vector $\mathbf{x}$ whose support is used as the label, and then generate $\mathbf{y}$ by simply applying (\ref{eq:CS_measurement}). Massive training data can be generated in this way, and the training process can be done in an off-line manner. Once parameters of BRNN is learned, using this neural network for inference, i.e., detecting active users in new dataset, is computational inexpensive, as it only involves a number of vector-matrix multiplications/summations and element-wise nonlinear operations.

\par
We would also like to emphasize that the required amount of training data depends on the number of nodes and the number of active nodes. With $K$ nodes in total, there are ${K\choose n}$ different labels for $n$ active nodes. This number increases quickly with the grow of $n$. Therefore, instead of generating the training data with a random active user number, we fix the number of active node close to the limit of traditional iterative algorithms for solving (\ref{eq:IHT}).

\section{Experiments}
In this section, we investigate the performance of the proposed BRNN for MUD in MTC. In the experiments, we consider $K=100$ users in total, and the active users transmit $N_\mathrm{s}$ pilot symbols with binary phase shift keying (BPSK) modulation. The channel is modeled by $L=6$ independent identically Rayleigh distributed taps. The receiver noise $\mathbf{n}$ is generated by a zero mean Gaussian vector with variance adjusted to have a desired value of the signal to noise ratio (SNR). As $K>\frac{N_\mathrm{s}+L-1}{L}$, the MUD problem is underdetermined.

\par
The proposed BRNN is compared with several iterative sparse estimation algorithms, including orthogonal matching pursuit (OMP)~\cite{6145475}, iterative hard thresholding (IHT)~\cite{blumensath2009iterative} and their extensions for block structure, i.e., BOMP~\cite{fu2014block} and BIHT~\cite{garg2011block}. The DNN proposed in~\cite{xin2016maximal} is also compared to emphasize the gain brought by BRNN. The detection of multiuser is considered to be successful if error occurs.

\par
In our experiments, we generate $8\times 10^6$ different samples for training, $10^5$ samples for verification and $10^5$ samples for testing. For all the generated data samples, we add additive white Gaussian noise with a signal-to-noise ratio (SNR) of 10 dB. In the training data and verification data, we randomly activate $6$ nodes, while in the test data, $n\leq 6$ active users are randomly selected. The optimizer adopted for training neural networks is stochastic gradient descent with a momentum 0.9 and a learning rate 0.01. The batch size is fixed to 250.

\subsection{Convergence Performance and Computation Efficiency}
\begin{figure}[!t]
\subfigure[Cross-entropy loss]{
\label{fig:conv1}
\includegraphics[width=4.2cm,height=3.2cm]{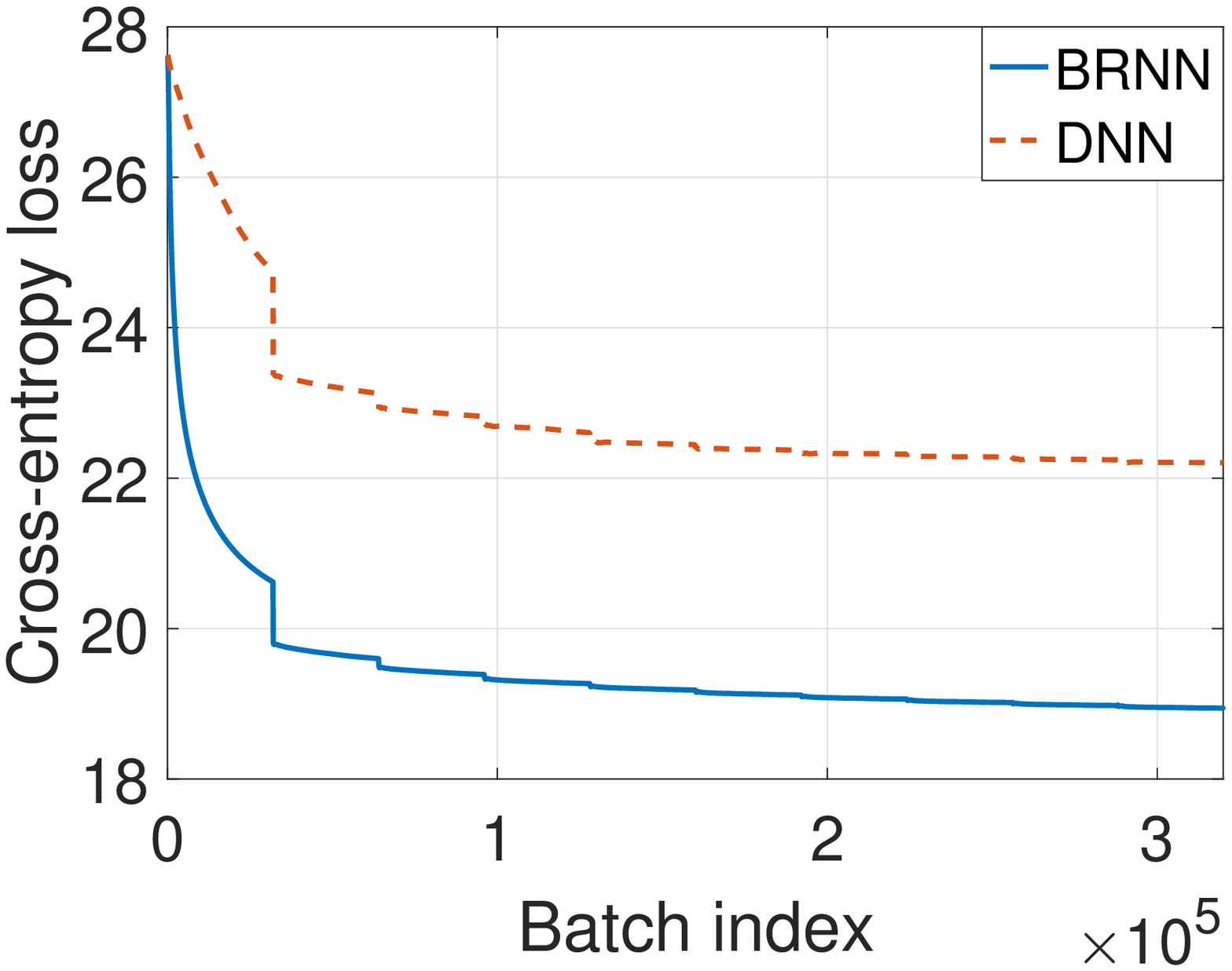}}
\subfigure[Ratio of successfully detected users]{
\label{fig:conv2}
\includegraphics[width=4.2cm,height=3.2cm]{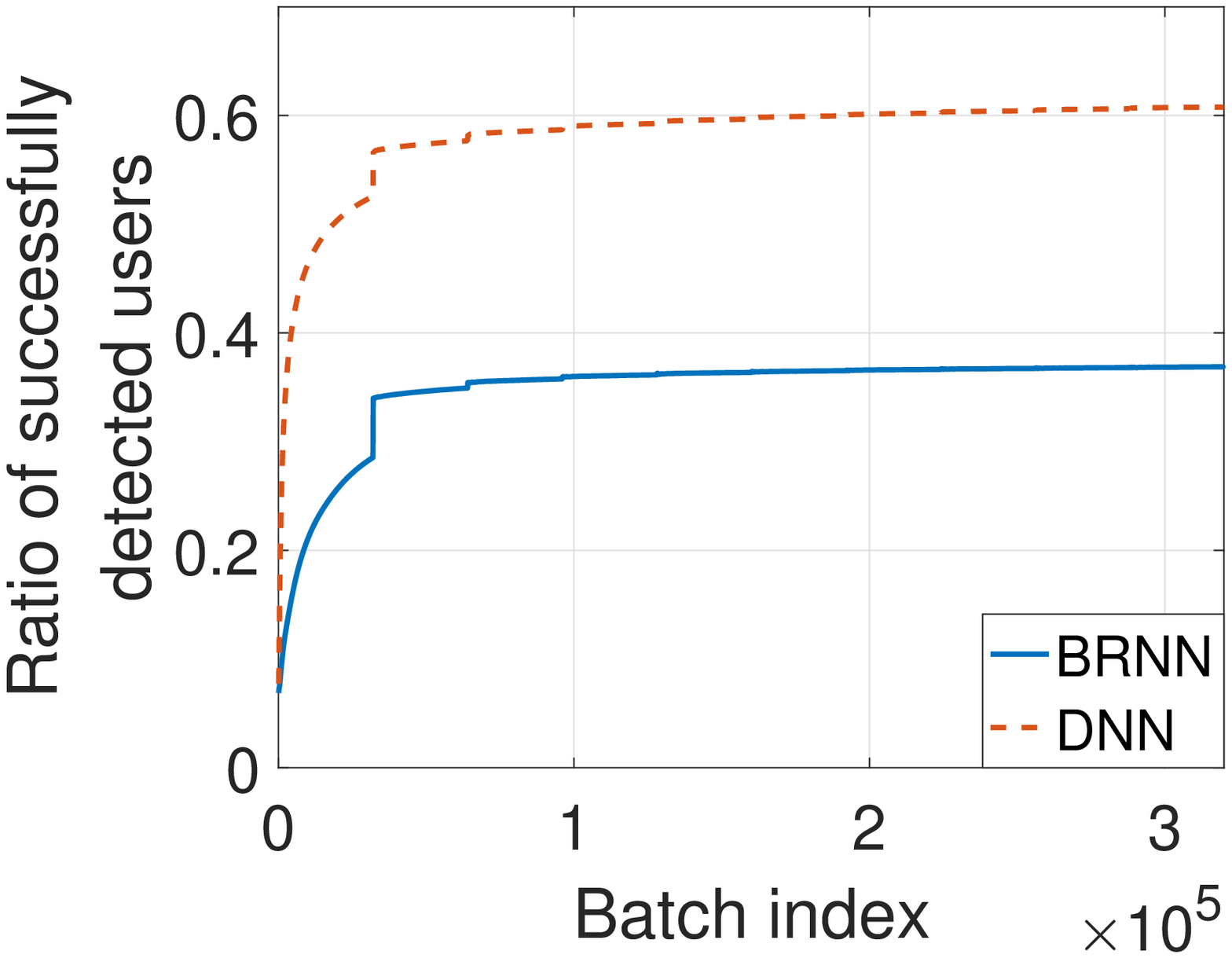}}
\caption{The convergence performance of DNN and BRNN.}
\label{fig:conv}
\end{figure}

We first investigate the convergence performance of DNN and the proposed BRNN in the training process. The pilot length of each user is fixed as $N_{s}=40$. We use the same initialization for DNN and BRNN as suggested in~\cite{he2015delving} and train the two neural networks with the same learning rate. The cross-entropy loss in the training process is calculated for every 100 batches, and the results are shown in Fig.~\ref{fig:conv1}. In addition to the cross-entropy loss, Fig.~\ref{fig:conv2} shows the ratio of successfully detected users in the training process. As shown in Fig.~\ref{fig:conv}, benefited from the block activation unit (\ref{eq:block_relu}), BRNN converges much faster and achieves a lower training loss than DNN, which fails to incorporate the prior knowledge on the structure of the signal support.

\par
Then we investigate the computation efficiency of inference using testing data. The averaged computing time for testing one data sample is given in Table \ref{time}. Note that although we use GPU to speed up the training process for BRNN and DNN, for a fair comparison of computational complexity we use CPU for the testing data for all the compared approaches including OMP, BOMP, IHT, BIHT, DNN and BRNN. These simulations are performed on a computer with a quad-core 4.2GHz CPU and 16 GB RAM, running under the Microsoft Windows 10 operating system. As shown in Table \ref{time}, for various settings of pilot length and user activation probability, deep learning approaches, i.e., DNN and BRNN, allow for more than $10$-fold decrease of the computing time. This significant improvement regarding to the computing complexity is owing to the fact that DNN and BRNN use a fixed number of matrix productions and nonlinear thresholding, while both OMP and BOMP involve computational complex matrix inverse operation, and both IHT and BIHT require a relatively large number of iterations to converge.

\begin{table}[]
\centering  
\caption{Averaged computing time for  multiuser detection (in seconds).}\label{time}
\begin{tabular}{|c|c|c|c|c|c|c|}  
\hline
$(N_{s},n)$ &DNN/BRNN  & IHT &BIHT & OMP &BOMP     \\ \hline\hline
$(30,3)$  &$2.57\times10^{-4}$ &$0.162$ &$0.173$ &$0.007$ &$0.007$\\  \hline
$(30,6)$  &$2.57\times10^{-4}$ &$0.009$ &$0.163$ &$0.007$ &$0.009$ \\  \hline
$(40,3)$  &$2.56\times10^{-4}$ &$0.183$ &$0.159$ &$0.009$ &$0.006$\\  \hline
$(40,6)$  &$2.56\times10^{-4}$ &$0.021$ &$0.177$ &$0.010$ &$0.010$\\  \hline
$(50,3)$  &$2.58\times10^{-4}$ &$0.193$ &$0.141$ &$0.011$ &$0.006$ \\  \hline
$(50,6)$  &$2.58\times10^{-4}$ &$0.085$ &$0.183$ &$0.0136$ &$0.012$\\  \hline
\end{tabular}
\end{table}

\subsection{Active User Detection Accuracy}
\begin{figure}[!t]
\subfigure[Success rate vs. $\frac{n}{K}$ ($N_{s}=40$)]{
\label{fig:sr1}
\includegraphics[width=4.2cm,height=3.2cm]{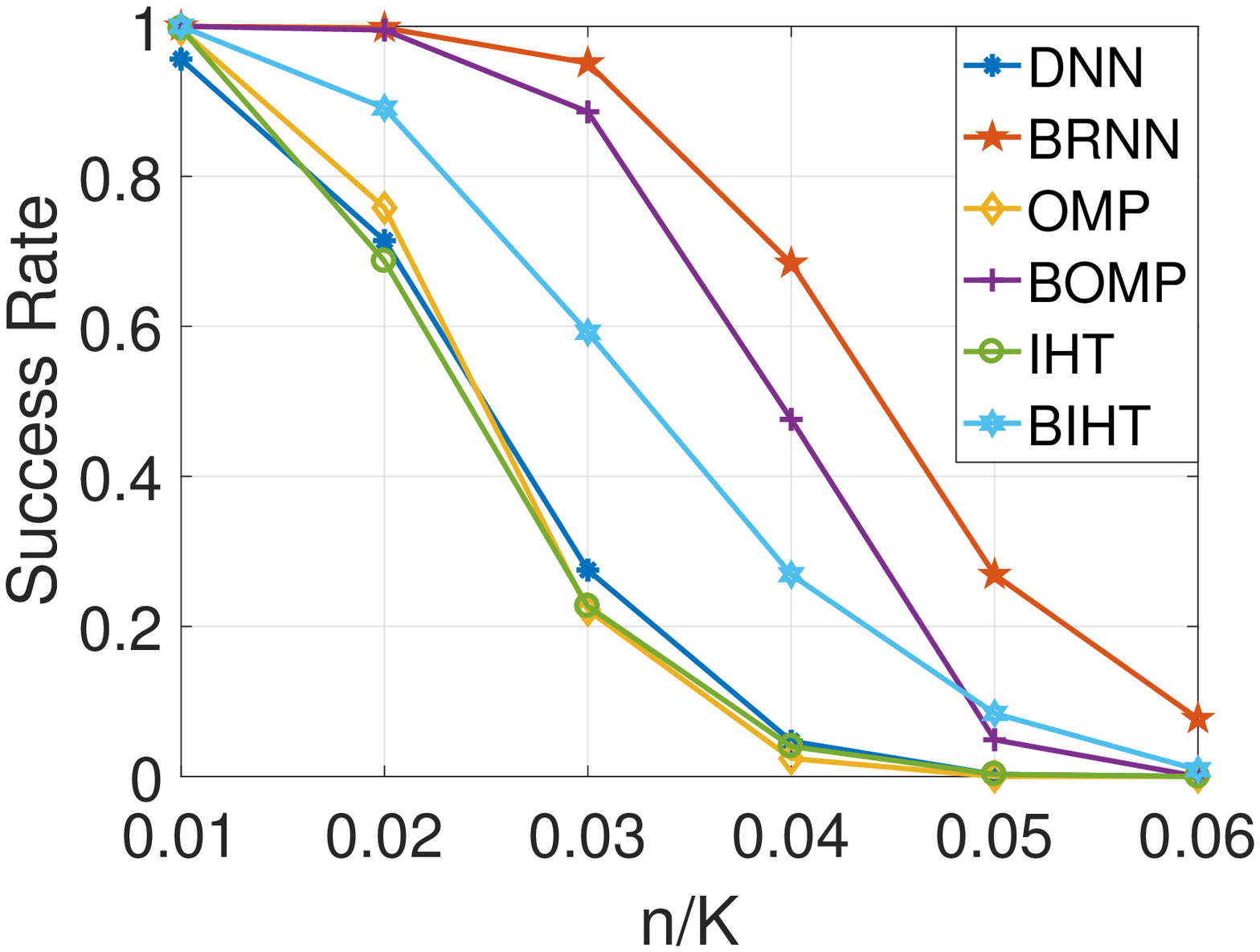}}
\subfigure[Success rate vs. $N_{s}$ ($n=4$)]{
\label{fig:sr2}
\includegraphics[width=4.2cm,height=3.2cm]{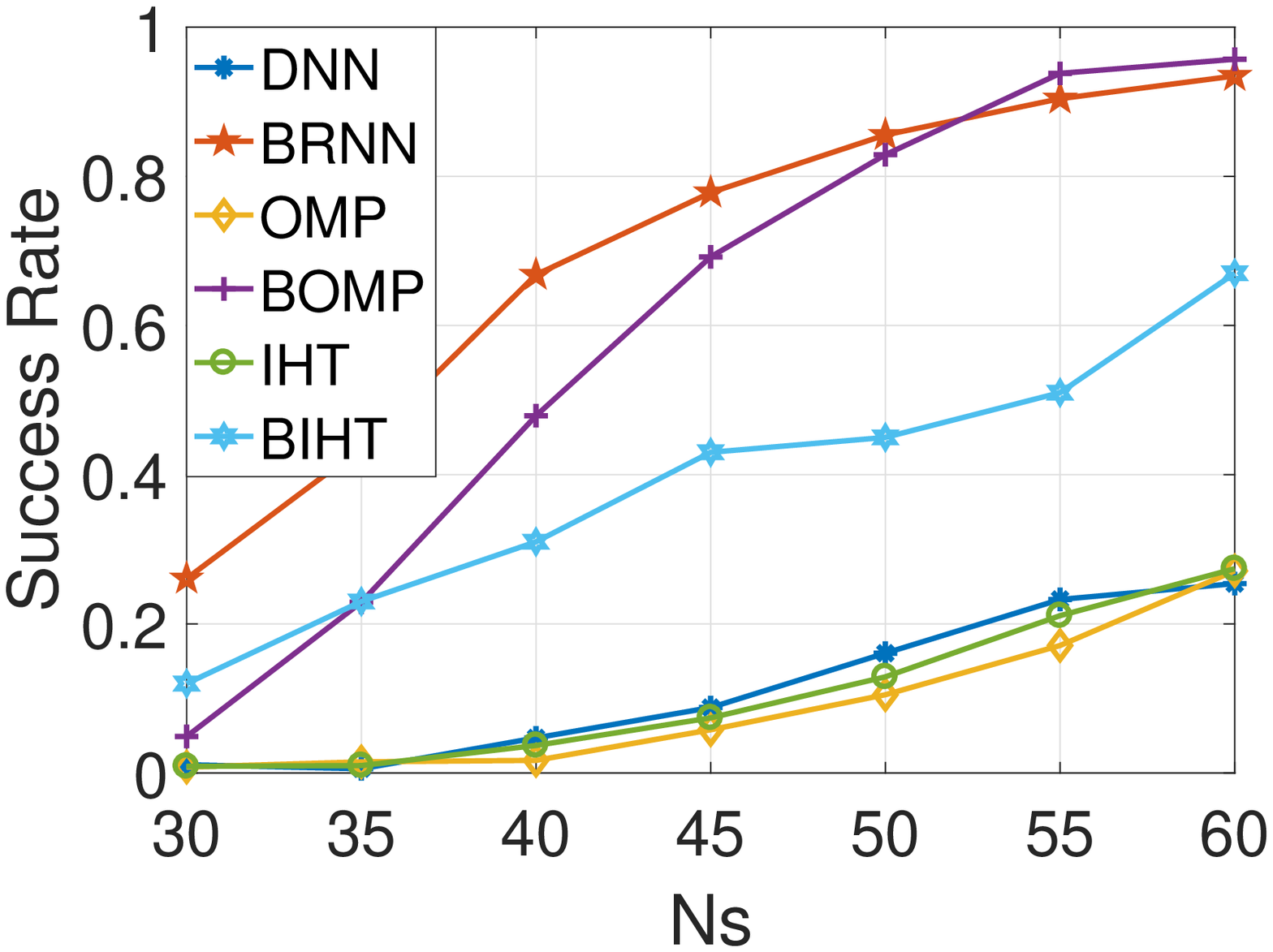}}
\caption{Performance of active users detection.}
\label{fig:sr}
\end{figure}

\par
In this experiment we study how the proposed approach performs with different numbers of active user and different lengths of pilot. In Fig.~\ref{fig:sr1}, the pilot length of each user is fixed as $N_{s}=40$. It is observed that the proposed BRNN achieves the highest active user detection success rate among the compared methods including OMP, BOMP, IHT, BIHT and DNN. Fig.~\ref{fig:sr2} shows the active user detection success rate with different pilot lengths, where the number of active users is set to be $4$. It is also observed that BRNN outperforms other approaches in most of the cases. Here, we would like to emphasize that there are various way to further improve performance of BRNN and DNN, e.g., using a larger size of training data and/or increasing the number of layers in the neural network, while the performance of OMP, BOMP, IHT and BIHT would not improve with more iterations.

\subsection{Channel Estimation Accuracy}

\begin{figure}[!t]
\subfigure[MSE vs. $\frac{n}{K}$ ($N_{s}=40$)]{
\label{fig:mse1}
\includegraphics[width=4.2cm,height=3.2cm]{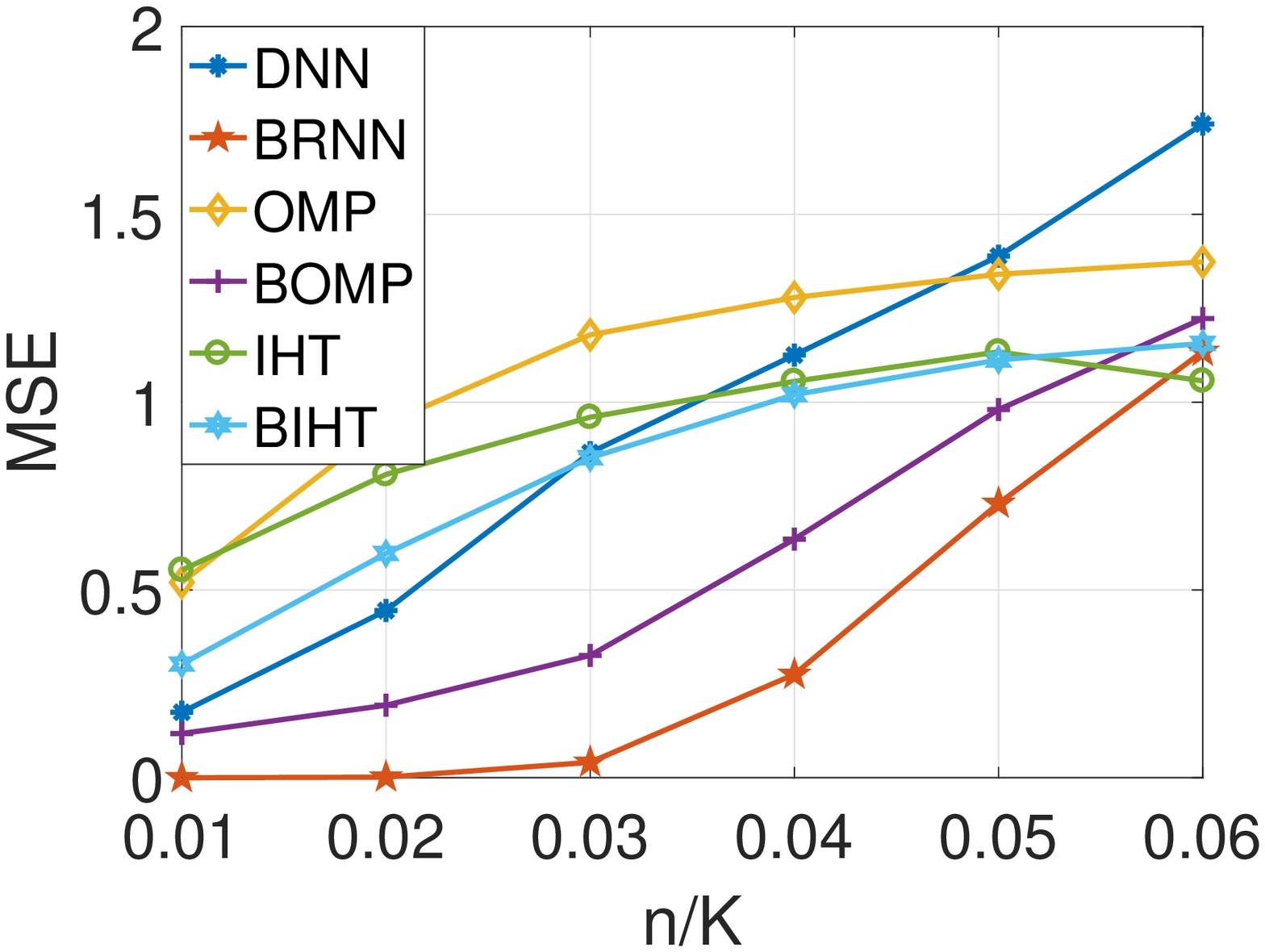}}
\subfigure[MSE vs. $N_{s}$ ($n=4$)]{
\label{fig:mse2}
\includegraphics[width=4.2cm,height=3.2cm]{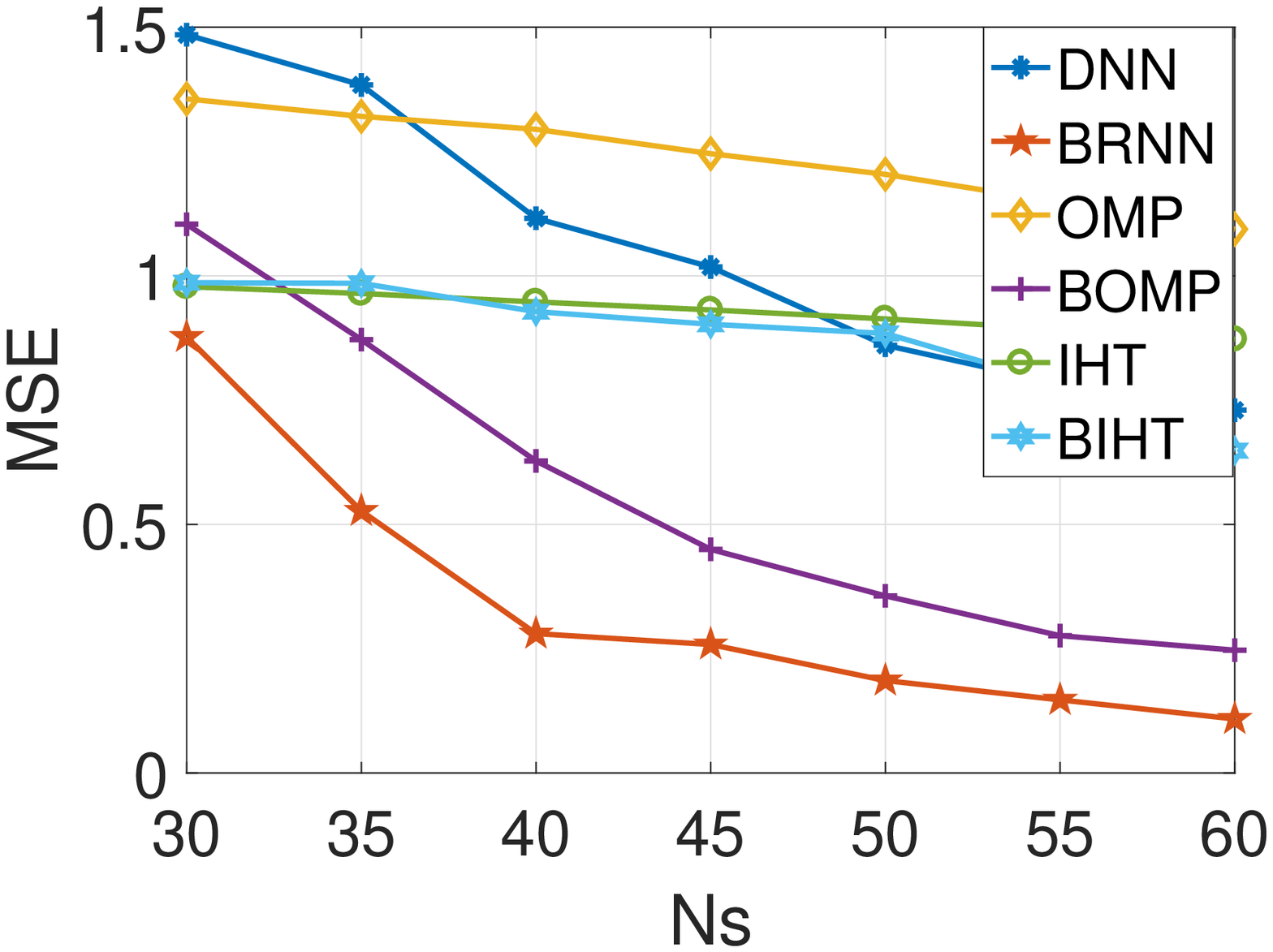}}
\caption{Performance of channel estimate.}
\label{fig:mse}
\end{figure}
In this experiment we show how does the proposed approach affect the channel estimation performance under different numbers of active user and different lengths of pilot. The channel is estimated by minimum mean square error estimator with the result of active user detect. In Fig.~\ref{fig:mse1}, the pilot length of each user is fixed as $N_{s}=40$. It is observed that the proposed BRNN achieves the smallest mean square error (MSE) among the compared methods including OMP, BOMP, IHT, BIHT and DNN. Fig.~\ref{fig:mse2} shows the MSE of channel estimation with different pilot lengths, where the number of active users is set to be $4$. It is observed that BRNN outperforms all the other approaches.

\section{Conclusion}
In this paper, we propose a novel deep neural network, called BRNN, for multiuser detection in massive MTC communication with with sporadically transmitted small packets and a low data rate. A new block activation layer is proposed in BRNN to capture the block sparse structure in the multiuser detection problem. In comparison with existing approaches, significant reduction of computing time and improvement of multiuser detection accuracy are achieved by the proposed approach.


%

%

%
%

\ifCLASSOPTIONcaptionsoff
  \newpage
\fi



%

\bibliographystyle{IEEEtran}
\bibliography{IEEEabrv,bib_paper}

%
%

%

%
%
%




\end{document}